\documentclass[useAMS,usenatbib] {mn2e}

\usepackage{times}
\usepackage{graphics,epsfig}
\usepackage{graphicx}
\usepackage{amsmath}
\usepackage{amssymb}

\newcommand{\msun}{\ensuremath{\rm M_{\odot}}}
\newcommand{\mb}  {\ensuremath{\rm M_{B}}}
\newcommand{\mhi} {\ensuremath{\rm M_{HI}}}
\newcommand{\dmhi} {\ensuremath{\rm dM_{HI}}}
\newcommand{\fnh} {\ensuremath{f({\rm N_{HI}})}}
\newcommand{\nh}  {\ensuremath{\rm N_{HI}}}
\newcommand{\lya}  {\ensuremath{{\rm Lyman}-\alpha}}
\newcommand{\dnh} {\ensuremath{\rm dN_{HI}}}

\newcommand{\acc}{\ensuremath{\rm atoms~cm^{-2}}}

\title {The HI column density distribution function in faint dwarf galaxies}
\author [Patra et al.]{
Narendra Nath Patra$^{1}$ \thanks {E-mail: narendra@ncra.tifr.res.in},
	Jayaram N. Chengalur$^{1}$ \thanks {E-mail: chengalu@ncra.tifr.res.in}, 
	Ayesha Begum$^{2}$  \thanks {E-mail: begum@astro.wisc.edu} \\ 
	$^{1}$ NCRA-TIFR, Post Bag 3, Ganeshkhind, Pune 411 007, India 
	$^{2}$ IISER-Bhopal,ITI Campus (Gas Rahat)Building,Govindpura, Bhopal - 23, India.
}
\date {}
\begin {document}
\maketitle
%\pagerange{\pageref{firstpage}--\pageref{lastpage}} \pubyear{}
\label{firstpage}

\begin{abstract}

  We present the HI column density distribution function,\fnh, as measured from dwarf galaxies observed as part of the Faint Irregular Galaxy GMRT (FIGGS) survey. We find that the shape of the dwarf galaxy \fnh\  is significantly different from the \fnh\ for high redshift Damped \lya\ absorbers (DLAs) or the \fnh\ for a representative sample of $z = 0$ gas rich galaxies. The dwarf \fnh\ falls much more steeply at high HI column densities as compared to the other determinations. While $\sim 10\%$ of the cross section above $\nh = 10^{20.3}~\acc$ at $z = 0$ is provided by dwarf galaxies, the fraction falls to $\lesssim 1\%$ by $\nh \sim 10^{21.5}~\acc.$ In the local universe, the contribution to the high \nh\ end of the \fnh\ distribution comes predominantly from the inclined disks of large galaxies. Dwarf galaxies, both because of their smaller scale lengths, and their larger intrinsic axial ratios do not produce large HI column densities even when viewed edge-on. If high column density DLAs/GRB hosts correspond to galaxies like the local dwarfs, this would require either that (i)~the absorption arises from merging and not isolated systems or (ii)~the observed lines of sight  are strongly biased towards high column density regions.

\end {abstract}

\section{Introduction}
\label{sec:int}

Neutral atomic gas at high redshifts can only be studied via the absorption lines that are produced when a gas cloud happens to lie in front of a more distant quasar. Much of what we know about the HI content of the early universe comes from the study of these absorption line systems. The absorption systems with the highest column densities ($\nh \gtrsim 10^{20.3}$~\acc), the so called Damped Lyman-$\alpha$ Absorbers (DLAs) are known to contain the bulk of the neutral gas at high redshifts. DLAs are of particular interest because at these column densities self shielding results in the ionised fraction of the gas being small \citep{wolfe05}.  As such, DLAs represent the most likely progenitors of the gas rich galaxies seen at the current epoch. 

Systematic surveys \citep[e.g.][]{wolfe86,prochaska09,noterdaeme09} have resulted in the discovery of a large number of DLAs. \citet{noterdaeme09} report the discovery of 937 DLAs using data from the SDSS~DR7 \citep{abazajian09}.  However, despite decades of study, the nature of DLAs remains unclear; in particular the size and morphology of the systems remains very poorly constrained by observations. Two extremes of the models that have been proposed are (i)~that DLAs arise from large rotating disks, much like the disks of modern day spirals \citep[e.g.][]{wolfe86,prochaska97} at the one end and (ii)~that DLAs arise from small systems analogous to the current day dwarf galaxies at the other \citep[e.g.][]{haehnelt98}. Recent numerical simulations suggest a more nuanced picture where a wide range of hosts gives rise to the observed DLAs \citep[see e.g.][]{pontzen08,cen12}.

Because the observable information available for DLAs is limited to the narrow pencil beam illuminated by the background quasar, there is limited scope for quantitative comparisons of the properties of DLAs and gas rich local galaxies. For DLAs with a compact background radio source, the spin temperature of the HI 21cm line can be measured \citep{wolfe79}. The spin temperature is typically higher than that observed in the disks of nearby spiral galaxies \citep[see e.g.][] {wolfe79,carilli96,kanekar03}, but similar to what might be expected in dwarf galaxies with low metallicity and low central pressures \citep{chengalur00,kanekar09}. Another direct observable from the absorption studies is the column density distribution function \fnh, which gives the expected number of absorbers with HI column density between \nh\ and \nh+\dnh\ per unit distance. For the local galaxy population also \fnh\ can be computed; this makes it  one of the few statistics which can be computed for both the high redshift as well as local populations. Further interest in studying \fnh\ comes from the idea that various physical processes, e.g. the onset of self-shielding, the threshold at which the gas goes from becoming dominantly atomic to dominantly molecular, would affect its shape \citep{altay11}. Studies of the evolution of \fnh\ could hence also lead to an understanding of the evolution with redshift of the physical conditions in neutral gas.

In this paper we present \fnh\ computed from the FIGGS survey \citep{begum08}. The FIGGS sample consists of extremely faint gas rich dwarfs; the median HI mass of the galaxies that we use here is only $\sim 1.4 \times 10^{7}$\msun. Our data set is complementary to the one used by \cite{zwaan05} to study \fnh ~at $z\sim 0$. Those authors used HI maps of 355 nearby galaxies from the WSRT based WHISP survey to determine \fnh. Their sample contains a representative mix of galaxy types, albeit being somewhat biased towards early types S0-Sb,
which they correct for using a type specific HI mass function. The \cite{zwaan05} study demonstrated that the cross-section for producing DLAs at $z\sim 0$ is dominated by large spiral galaxies. The \fnh\ that they compute is hence essentially that corresponding to large galaxies. Our current sample excludes large spirals, and the \fnh\ that is computed here corresponds to that in the smallest star forming galaxies. There are a number of reasons why it is interesting to study the \fnh\ in such small galaxies. Firstly, since the gas in dwarf galaxies is dust poor, one would expect the transition from atomic to molecular gas to happen at a higher column density than in spirals \citep{mckee10,welty12}. Further, unlike in spiral galaxies the gas in dwarfs is not in a thin, dynamically cold disk \citep{roychowdhury10}. For these reasons, it is not apriori obvious that the \fnh ~in dwarf galaxies would be similar to that in large spirals. Finally, in hierarchical models of galaxy formation, small objects form first, and one would expect that the properties of the gas rich objects in the early universe would more closely resemble dwarfs, than large spirals. These are the themes that we explore in the rest of this paper.

\section{Description of the Sample and primary data }
\label{sec:sample}

The data that we use were gathered as part of the FIGGS survey \citep{begum08}. The survey used a sample of 65 dwarf galaxies, selected from the \citet{kk04} catalog, that satisfied the following selection criteria: (i)~absolute blue magnitude $M_B \ge -14.5$~mag, (ii)~ HI integrated flux $> 1.0$ Jy kms$^{-1}$, and (iii)~optical B-band major-axis $\ge 1.0$ arcmin.
Three of the galaxies in the FIGGS sample were not detected in HI. One galaxy had a companion that was detected in HI, which we also use for our analysis. This leaves us with a total of 63 galaxies from the FIGGS survey. A further 16 fainter local galaxies which satisfy the following selection criteria
(i)~absolute blue magnitude,  M$_{B}$ $\ge$  --12.0~mag, and (ii)~integrated HI flux $\ge$  0.5 Jy kms$^{-1}$ were later added to the sample \citep{patra12}.

\begin{figure*}
\begin{center}
\begin{tabular}{cc}
\resizebox{75mm}{!}{\includegraphics{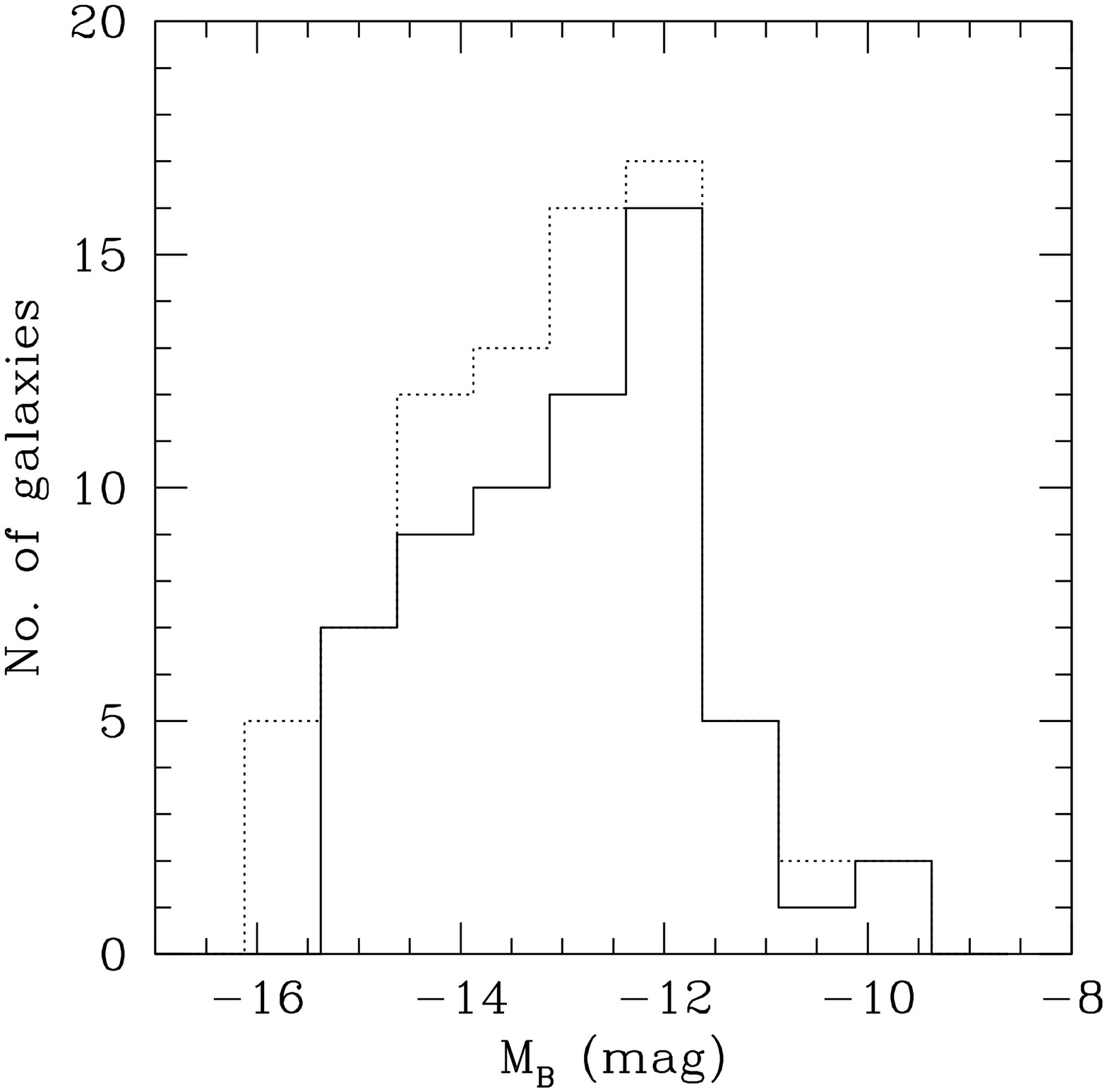}} &
\resizebox{75mm}{!}{\includegraphics{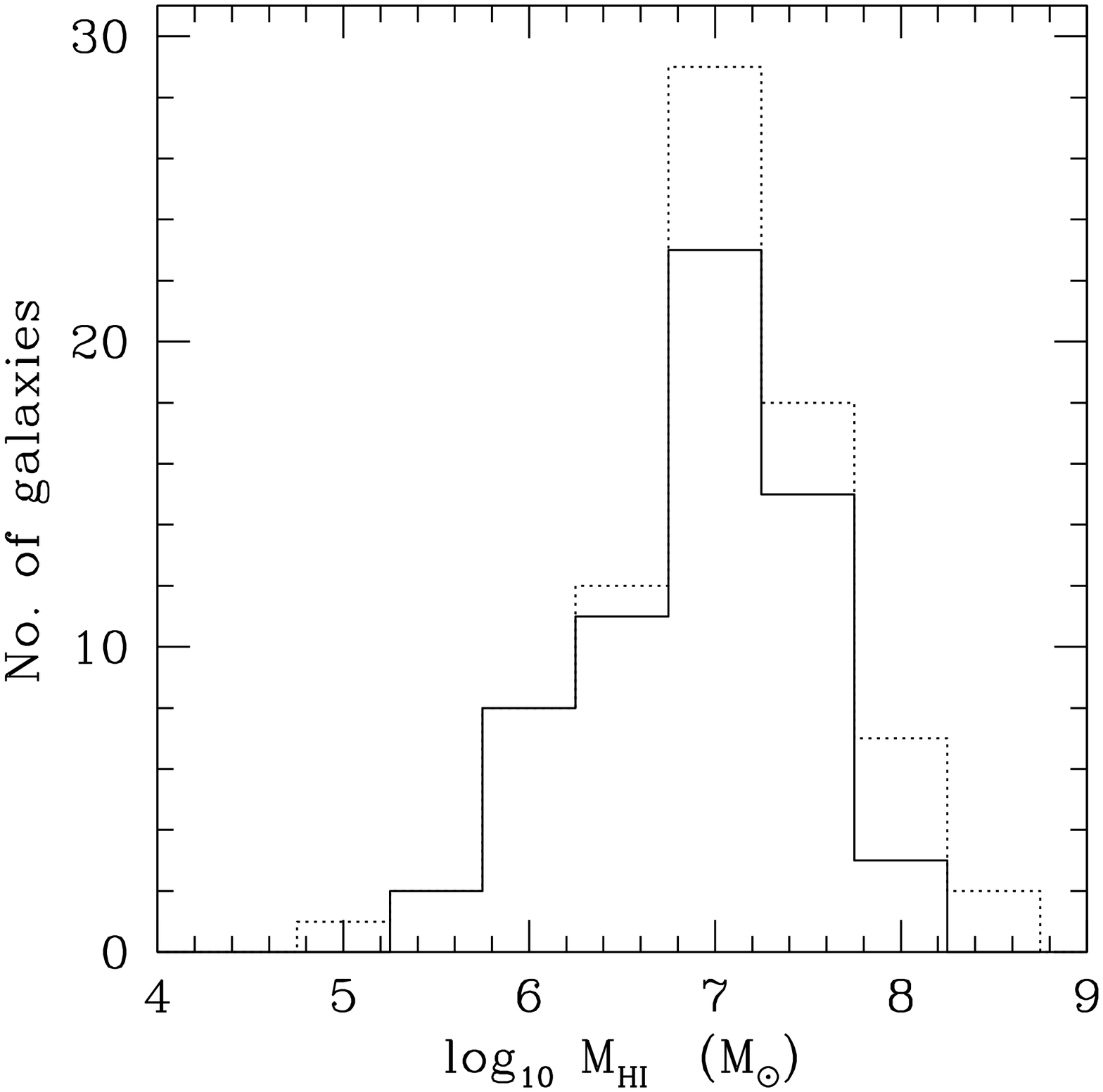}} \\
\end{tabular}
\end{center}
\caption{ Distribution of absolute blue magnitude \mb\ (left panel) and \mhi\ (right panel). Because of problems with the high resolution images of some galaxies, we use only a subset of the total sample for the analysis in this paper (See the text for more details). The dotted lines represent the entire FIGGS sample, whereas the solid lines are for the galaxies whose data are used in this paper. }

\label{fig:samp}
\end{figure*}

The GMRT has a hybrid configuration \citep{swarup91}, i.e. with a mix of short and long baselines. Consequently, images at a range of resolutions can be made from a single GMRT observation. For the FIGGS galaxies we have data cubes and moment maps at resolutions ranging from $\sim 40^{''}$ to $\sim 4^{''}$, corresponding to linear resolutions of 850~pc to 85~pc at the median distance to the galaxies in the sample. For this study we use only those galaxies for which good quality images were available at all resolutions. This leaves us with a total of 62 galaxies. Figure~\ref{fig:samp} shows the distribution of absolute blue magnitude and HI mass for our sample galaxies. For this sub-sample, the median HI mass is $\sim 1.4\times 10^7$\msun, the median blue magnitude is $\mb \sim -12.3$~mag, and the median distance is $\sim 4.4$~Mpc.

\begin{figure*}
\begin{center}
\epsfig{file=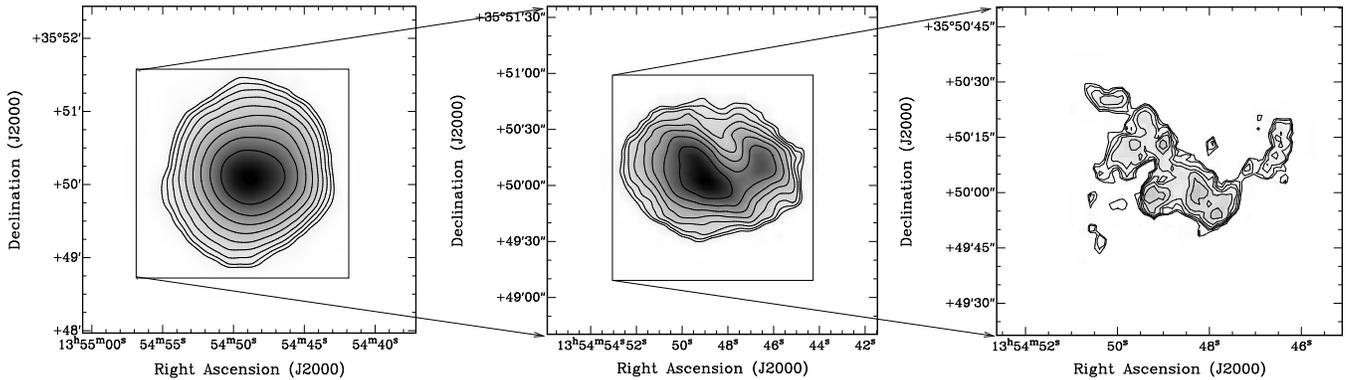,width=7in}
\end{center}
\caption{ Integrated HI maps for UGC~8833 ($\mhi = 10^{7.2} \msun$) at 
three different resolutions, viz. $\sim 40^{''}$ (left panel), 
$\sim 12^{''}$ (middle panel) and $\sim 4^{''}$ (right panel). In all 
three panels the contours spacing is $\sqrt{2}$. The first contour is at 
$6, 10 ~\& ~30 \times 10^{19}$ \acc respectively for the left, middle
and right panels.}
\label{fig:contour}
\end{figure*}

In Fig.~\ref{fig:contour} we show the integrated HI image for a representative sample galaxy, UGC~8833, which has an HI mass of $10^{7.2}$\msun\  and an inclination $\sim 28^{o}$. Maps at three different resolutions, viz. $\sim 40^{''}$, $12^{''}$  and $4^{''}$ are shown. As can be clearly seen, the low resolution images smooth over the high column density clumps that are seen at higher resolutions. Conversely, the smooth diffuse emission that is seen in the low resolution images is largely resolved out in the high resolution images. Clearly, for any individual galaxy one would need to use data from a range of resolutions to get an accurate estimate of the HI column density distribution. We return to this issue in the next section where we compute the HI column density distribution for our sample of galaxies.

\section{The Column Density Distribution Function}
\label{sec:fnhi}

The HI column density distribution \fnh\ was first used in the context of gas seen in absorption against distant quasars. It is  defined such that \fnh\dnh dX gives the number of systems with gas column density between \nh\ and $\nh + \dnh$ that would be encountered within an absorption distance dX \citep{bahcall69}. The quantity \fnh\ is directly computable from absorption spectra \citep[see e.g.][]{wolfe86, prochaska09, noterdaeme09} and is hence fairly well known at redshifts $\gtrsim 2$. At redshifts below $\sim 2$ the \lya\ line cannot be observed from ground based telescopes; \fnh\ is consequently not well measured at these redshifts. At the lowest redshifts (i.e. $z \sim 0$) however, \fnh\ can be computed from resolved images of the HI disks of galaxies and knowledge of the HI mass function \citep[see e.g.][]{rao93,zwaan05}. Specifically,
from the integrated HI map of a galaxy one can compute the total area in the map that is covered by gas with column densities between \nh\ and $\nh + \dnh$. This area is clearly dependent on the inclination of the galaxy to the line of sight; however if one has a large sample of galaxies with random inclinations to the line of sight, the average area $A(\nh,\mhi)$, provided by galaxies in some HI mass bin would automatically represent the average over different inclinations. In order to compute the average area along a random line of sight through the universe, this average area has to be further normalised by the number density of galaxies with HI mass in the given mass bin, i.e. the HI mass function. Putting this all together, (and also using the fact that at $z = 0$, dX/dz = 1) \fnh\ can be computed as \citep[see e.g.][]{zwaan05}.

\begin{equation}
\fnh d\nh = {\frac{c}{H_0}} \int \psi(\mhi) A(\nh,\mhi) \dmhi
\label{eqn:fnhi}
\end{equation}

where 
\begin{equation}
\psi(\mhi) = \frac{\phi(\mhi)}{ln(10.0) \mhi}
\end{equation}

and $\phi(\mhi)$ is the usual HI mass function (per unit interval of $\log10(\mhi)$) which is generally parametrised as a Schechter function. Here we use the Schechter function parametrisation provided by \citet{martin10} with $\alpha = -1.33, \phi_{*} = 4.8\times 10 ^{-3} $ Mpc$^{-3} \rm{dex}^{-1}$, ${\mhi}_{*} = 10^{9.96}$\msun. The integral in Eqn.~(\ref{eqn:fnhi}) was computed by summing the integrand computed over logarithmic bins in \mhi, the bin width was taken to be 0.3. We have confirmed that the result is not very sensitive to the chosen bin width, results obtained using bin widths
of 0.2 and 0.4 overlap within the error bars. The slope of the faint end of the HI mass function remains somewhat uncertain. In the tabulation in  \citet{martin10} of various determinations of the HI mass function, the values reported for the faint end slope in different studies vary from $-1.20$ to $-1.41$. We conservatively take the error bars to be the quadrature sum of the variation in \fnh\ that one gets using these extreme values of the faint end slope and the errors obtained from bootstrap re-sampling over 100 runs.

\begin{figure*}
\begin{center}
\begin{tabular}{cc}
\resizebox{80mm}{!}{\includegraphics{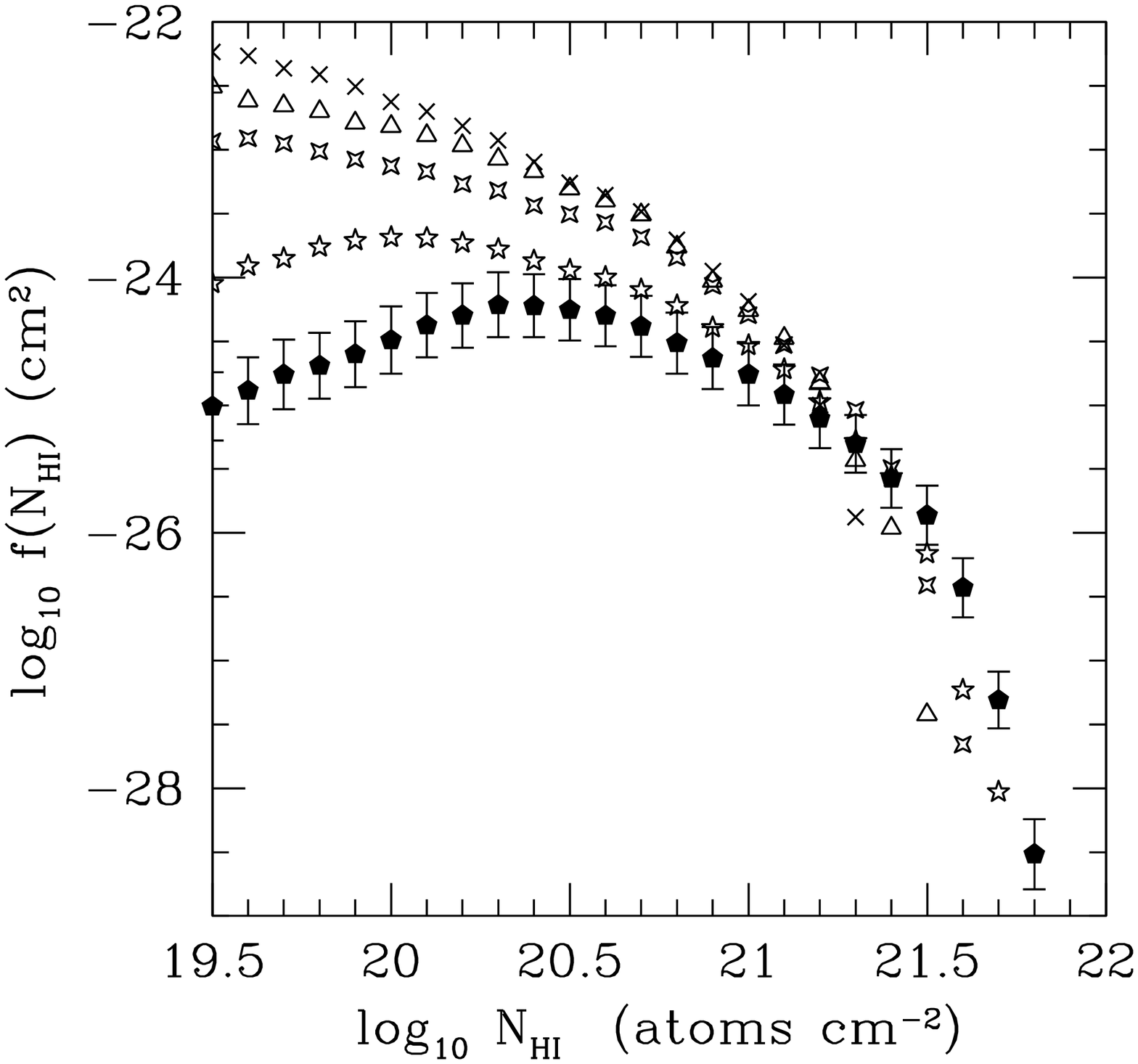}} &
\resizebox{80mm}{!}{\includegraphics{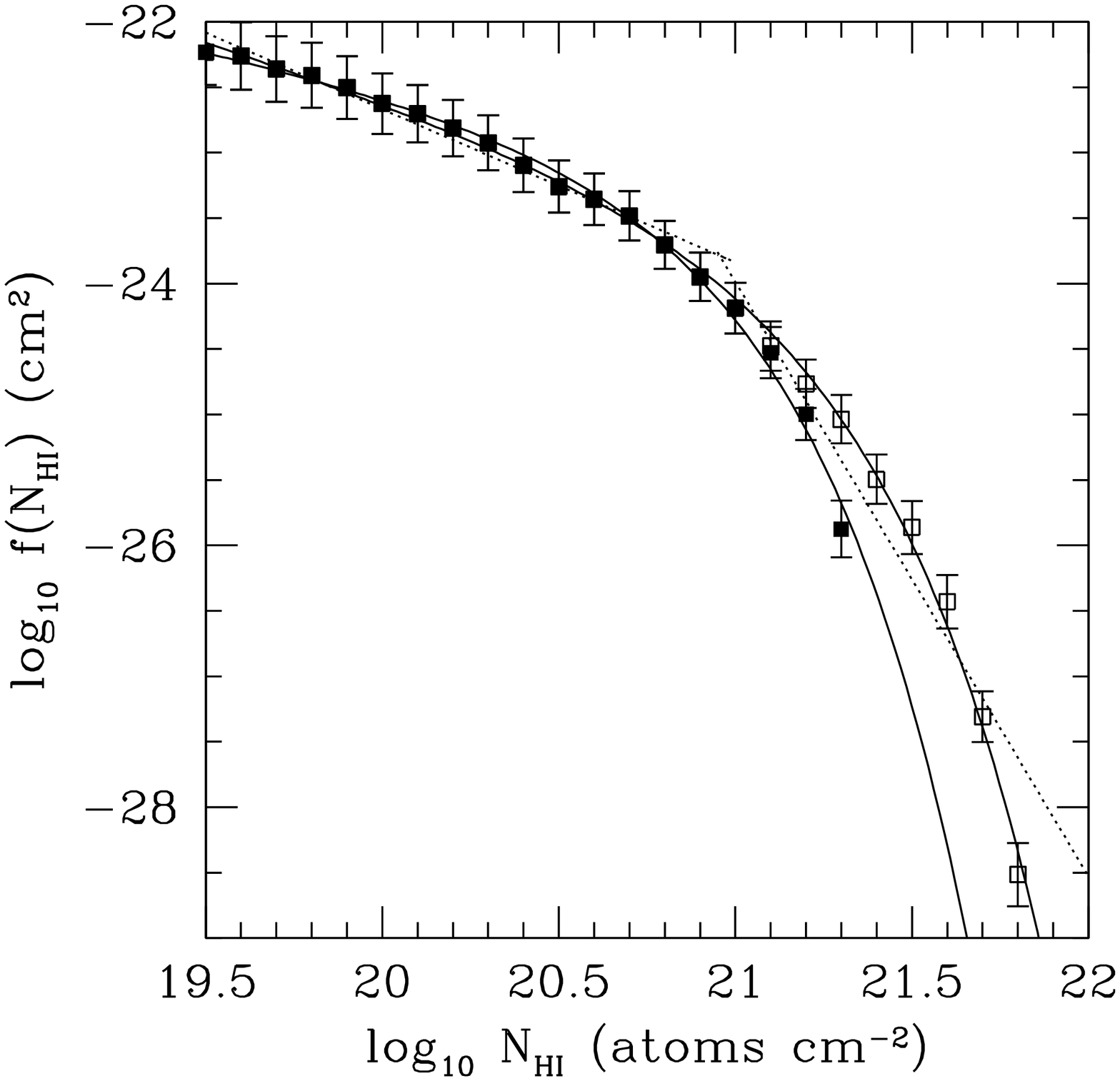}} \\
\end{tabular}
\end{center}
\caption{Left panel: The HI column density distribution function \fnh\ computed
from images at different resolutions. The crosses use the $40^{''}$ resolution data, the triangles are for the $25^{''}$ resolution data, the hollow four pointed stars are for the $12^{''}$ resolution data, the hollow five pointed stars are for the $6^{''}$ resolution data and the filled pentagons are for the $4^{''}$ resolution data. For clarity error bars are shown only for the $4^{''}$ data (for which the  error bars are largest). Right panel: The hybrid \fnh\ (hollow squares) and the \fnh\ derived from the $40^{''}$ resolution data (filled squares). Gamma function fits (solid line) are shown for both sets of data.  For both data sets, the break point for a piece wise linear fit is at $\log(\nh) \sim 21.0$, for clarity this is only shown for the hybrid curve (dotted lines).}
\label{fig:fnhi_res}
\end{figure*}

Fig.~\ref{fig:fnhi_res} (left panel) shows \fnh\ computed from images at different resolutions. As discussed above, the \fnh\ computed from the low resolution images will underestimate the true \fnh\ at high column densities, while the \fnh\ computed from the high resolution images will underestimate \fnh\ at low column densities. This is clearly seen in the Fig.~\ref{fig:fnhi_res} (left panel). We hence also use a ``hybrid'' \fnh. For the hybrid \fnh, the value at any \nh\ is set to the maximum value obtained for the different resolutions. We note that the hybrid curve differs from the curve made using the low resolution $40^{''}$data only at the highest column densities, viz. $\nh \gtrsim 10^{21.5}$\acc. \cite{zwaan05} point out that a hybrid \fnh\ computed in this way will be an over-estimate, since the gas that is seen at high column densities in the high resolution images will also, after the smoothing that happens in the low resolution image, contribute to gas at lower column densities. In plots below we hence show results computed from both the hybrid \fnh\ and the \fnh\ computed from the $40^{''}$ resolution images. 

\begin{figure}
\epsfig{file=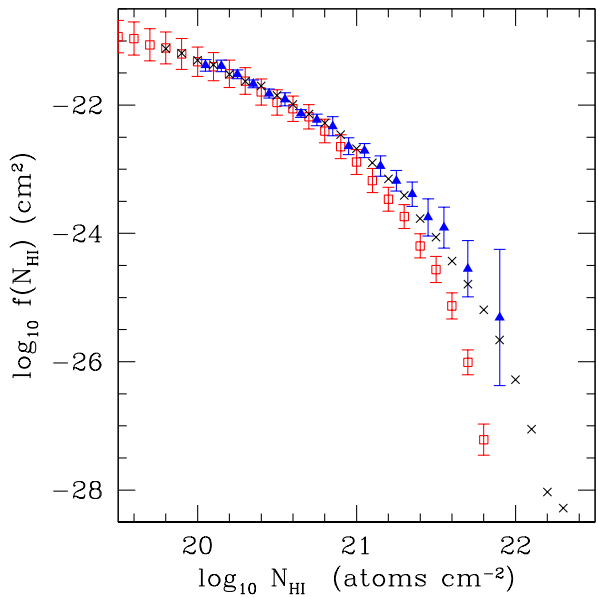,width=80mm}
\caption{Comparison of the shape of \fnh\ computed from dwarf galaxies (hollow squares) with the \fnh\ for DLAs (\citet{noterdaeme09}, filled triangles) and for the $z=0$ galaxy population (\citet{zwaan05}, crosses). As can be seen the \fnh\ for dwarfs falls sharply at high column densities as compared to the other two curves.}
\label{fig:dla_cmp}
\end{figure}

\subsection{Results and Discussion}
\label{ssec:res}

Fig.~\ref{fig:fnhi_res} (right panel) shows the hybrid \fnh, as well as the \fnh\ derived from the $40^{''}$ resolution data. To allow easy comparison with previous works, we also show two different fits, i.e. a broken power law, and a gamma function defined as:

$$
f\left(N_{HI}\right) = \alpha\left(\frac{N_{HI}}{N_\ast}\right)^{-\beta} e^{-\left(N_{HI}/N_\ast\right)}
$$
\\

 Below (see also Tab~\ref{tab:gamma}) we compare the \fnh\ computed here with those determined by \cite{zwaan05} and \cite{noterdaeme09}.
In the piece wise power law fit to the dwarf galaxy \fnh\ the break point which gives the best fit is at $\log(\nh) \cong 21.0$ for the hybrid \fnh, and $20.9$ for the \fnh\ derived from the $40^{''}$ resolution data. As can be seen from Fig.~\ref{fig:fnhi_res} (right panel) \fnh\ is curved at high column densities and the gamma function provides a much better fit to the data than a broken power law. This curvature of the \fnh\ curve is very similar to what \cite{zwaan05} found for their WHISP sample. 

\begin{table*}
\begin{tabular}{|l|c|c|c|c||c|}
\hline
Parameter&\cite{zwaan05} &\cite{noterdaeme09} &FIGGS&FIGGS\\      
&&&(Hybrid)&$40^{''}$\\
\hline       
$\log(\alpha)$ &-22.91&-22.75&-23.29 $\pm$ 0.04& -22.79 $\pm$ 0.06\\             
$\beta$  &1.24        &1.27          & $0.88 \pm 0.04$ & $0.57 \pm 0.07$\\                                   
$\log{N_\ast}$  &21.2 &21.26         & $20.82 \pm 0.01$ & $20.55 \pm 0.03$\\                                
\hline
\end{tabular}
\label{tab:gamma}
\caption{Parameters of the Gamma function fit to \fnh.}
\end{table*}

A particularly interesting result to come out of the earlier data sets is that the shape of the \fnh\ derived for the local galaxy population agrees well with that derived from DLAs \citep{zwaan05,prochaska09}. This is a surprising result for a number of reasons, viz. (i)~the total neutral gas density evolves, changing by about a factor of 2 between high redshifts and $z \sim 0$ \citep{zwaan03,noterdaeme09}, (ii)~the morphological mix of galaxies is significantly different at high redshifts \citep{conselice05}, and  (iii) DLAs have significantly lower metallicity than low redshift spiral galaxies that dominate the HI cross section at $z=0$. The non-evolution of \fnh\ has led to suggestions that it is set by some universal process that shapes the gas distribution \citep{erkal12}. A possible explanation is geometric -- for gas in a thin disk, averaging over different orientations would lead to an \fnh\ that falls as $\nh^{-3}$ at column densities larger than the maximum face on column density of the disk \citep{milgrom88,erkal12}.

In Fig.~\ref{fig:dla_cmp} we show the \fnh\ computed by \cite{zwaan05}, \cite{noterdaeme09} and for the FIGGS sample. All the \fnh\ curves have been normalised to have the same value at $N_{HI} = 10^{20}$~\acc\ so that one is comparing only the difference in shapes. As can be seen the dwarf galaxy \fnh\ differs markedly from the other two in that it falls much more steeply at the high column density end ($\nh \gtrsim 10^{21}$\acc). The HI cross section at $z \sim 0$ is dominated by large galaxies \citep{zwaan05}, so the difference between the FIGGS \fnh\ and that determined by \citet{zwaan05} is essentially the difference between large galaxies and dwarfs. It is interesting that the \fnh\ for dwarf galaxies begins to fall below  the other two curves for  column densities larger than $\log(\nh) \sim 21.0$~\acc. For the \fnh\ derived from the THINGS sample \cite{erkal12} also find a turnover at $\nh \sim 10^{21}$~\acc. They attribute this feature to the HI cross section above this column density being dominated by highly inclined thin disks. The high column densities come from the long path lengths that the line of sight traverses through these disks. The sharp decline of the dwarf galaxy \fnh\ at high \nh\ is presumably because in dwarfs the HI gas is not in a thin disk. \citet{roychowdhury10} show that for the FIGGS sample, the median intrinsic axial ratio is $\sim 0.6$.  In this context it is worth noting that the fall off at high \nh\ is found in the dwarf data despite the relatively high linear resolution of $\sim 100$pc of our data (In contrast, the \cite{zwaan05} sample was observed with a resolution $\sim 1.3$kpc).  In DLAs the drop in \fnh\ at $\nh\sim 10^{22}$\acc\ is attributed to the onset of $H_2$ formation; above this column density, the gas transitions to being mostly molecular \citep[see e.g][]{altay11}. Since this transition happens at still higher column densities in low metallicity dwarfs \citep{mckee10,welty12}, it cannot be the cause of the drop off that we see in the dwarf \fnh. 

We note that the \fnh\ computed here as well as in \citet{zwaan05} assumes that the emission is optically thin. \citet{braun12} models the HI emission as arising from an isothermal gas distribution and finds that
in this model optical depth corrections could lead to HI column densities as high as few times $10^{23}$~\acc. This is in contrast to the current understanding that the gas becomes predominantly molecular
at these column densities \citep[e.g.][]{schaye01}. 
If optical depth corrections are indeed important, the comparison of the \fnh\ of dwarfs with DLAs may need some revision at the high column density end. However, the comparison of \fnh\ between the dwarf \fnh\ and the \cite{zwaan05} result should remain largely unaffected,
since both will have similar corrections. 

\begin{figure}
\epsfig{file=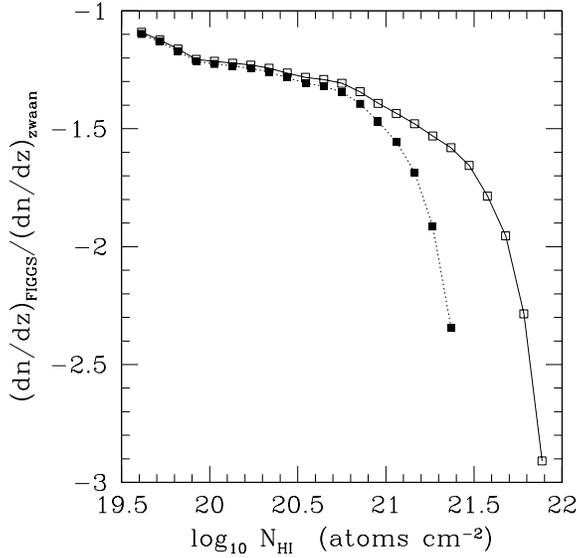,width=3in}
\caption{Ratio of $\frac{dn}{dz}$  computed from dwarf galaxies with that computed for the $z=0$ galaxy population \citep{zwaan05}. The solid squares are for the \fnh\ derived from the $40^{''}$ resolution data while the hollow squares are for the hybrid \fnh. }
\label{fig:dndz_ratio}
\end{figure}

Fig.~\ref{fig:dndz_ratio} shows the ratio  $dn/dz(\nh)$ as computed from the FIGGS data and from the \citet{zwaan05} sample. $dn/dz(\nh)$, is  the integral over \dnh\ of \fnh, and gives the average number of absorbers with column density greater than \nh\ per unit redshift. As can be seen, above the DLA column density limit ($\nh = 10^{20.3}$~\acc) dwarfs (with $\mhi \lesssim 10^{8}$\msun) contribute about 10\% to the cross-section. However the fraction drops rapidly with increasing \nh, and by $\nh \sim 10^{21.5} \acc$, dwarfs contribute less than 1\% to the cross-section. In a model in which the host galaxies of DLAs were similar to the $z=0$ dwarfs one could match the DLA number count (i.e. $dn/dz(N_{HI}=10^{20.3})$) by scaling up the number density of dwarfs at high redshift. However, such a model would predict $\lesssim 3$ DLAs with $\nh > 10^{21.5}$\acc\ in the SDSS volume, instead of the $\gtrsim 30$ found by \citet{noterdaeme09}. Further, DLAs with $\nh \sim 10^{22}$\acc\ have been found in the SDSS survey \citep{noterdaeme09,kulkarni12,noterdaeme12}. From the \fnh\ for DLAs, \citet{noterdaeme09} estimate that only $\sim 1$ such high column density DLA should be detected in the SDSS survey volume. Since the \fnh\ for dwarfs falls sharply at high column densities, the probability of finding two such high column density absorbers  in the SDSS volume is vanishingly small for a host population dominated by dwarfs. Similarly high column densities are also seen in the DLAs arising from the host galaxies of high redshift GRBs (GRB-DLAs). Photometry of GRB hosts suggests that they are small galaxies \citep{chen09}, albeit somewhat brighter than the dwarfs considered here. Nonetheless, the sharp fall off in \fnh\ at high column densities for dwarfs strongly supports the suggestions \citep[see e.g.][]{prochaska07} that GRBs probe biased regions of the ISM, i.e. those with the highest column density. In the case of ordinary DLAs, as argued above, models where the host galaxies are all similar to $z=0$ dwarfs appear to be ruled out. However models with a larger fraction of dwarfs but either (i)~a sufficient number of disk galaxies to provide the observed cross section at high column densities, and/or (ii)~where a signficant fraction of the high column density gas arises from merging dwarfs may still be consistent with both the observed DLA \fnh\  and the large velocity spreads observed in the DLAs low ionisation metal lines \citep[see e.g.][]{prochaska97}.

To summarise, we determine the \fnh\ in faint local dwarfs, and find that it falls off significantly faster at high column densities than the \fnh\ in DLAs or in the \fnh\ determined from a representative sample of the local galaxy population. For the local galaxy population, at the high \nh\ end the dominant contribution to \fnh\ comes from the disks of large spiral galaxies viewed edge on. Isolated galaxies like the $z=0$ dwarfs hence cannot form the dominant host population of high redshift DLAs and GRB-DLAs unless there is significant biasing of the observed lines of sight and/or corrections for the opacity of the HI line is
important even for dwarf galaxies. The FIGGS sample
is one of the largest existing samples of faint dwarf galaxies with
high resolution HI images. We hence expect that further significant
progress on the shape of \fnh\ in dwarfs will require data from HI surveys 
using the next generation telescopes, i.e. the Australian SKA 
Pathfinder (ASKAP; Deboer et al. 2009), MeerKAT \citep{jonas09} in South 
Africa and APERTIF \citep{verheijen08} in the Netherlands.

Discussions with and comments from Nissim Kanekar as well as very useful comments from the anonymous referee are gratefully acknowledged.


\begin{thebibliography}{}

\bibitem[\protect\citeauthoryear{Altay et al.}{2011}] {altay11} 
Altay G., Theuns T., Schaye J., Crighton N.~H.~M., Dalla Vecchia C., 2011, 
ApJ, 737, L37 

\bibitem[\protect\citeauthoryear {Abazajian et al.} {2009}]{abazajian09}
 Abazajian, K.~N., Adelman-McCarthy, J.~K., Ag{\"u}eros, M.~A., et al.\ 2009, ApJ, 182, 543 

\bibitem[\protect\citeauthoryear{Braun}{2012}]{braun12} Braun R., 2012, ApJ, 749, 87 

\bibitem[\protect\citeauthoryear{Bahcall \& Peebles}{1969}] {bahcall69} Bahcall J.~N., Peebles P.~J.~E., 1969, ApJ, 156, L7 

\bibitem[\protect\citeauthoryear{Begum et al.}{2008}] {begum08} 
Begum A., Chengalur J.~N., Karachentsev I.~D., Sharina M.~E., Kaisin S.~S., 2008, MNRAS, 386, 1667 

\bibitem[\protect\citeauthoryear{Carilli et al.}{1996}] {carilli96} Carilli C.~L., Lane W., de Bruyn A.~G., Braun R., Miley G.~K., 1996, AJ, 111, 1830 

\bibitem[\protect\citeauthoryear{Cen}{2012}] {cen12} Cen R., 2012, ApJ, 748, 121 

\bibitem[\protect\citeauthoryear{Chen et al.}{2009}] {chen09} 
  Chen H.-W., et al., 2009, ApJ, 691, 152 

\bibitem[\protect\citeauthoryear{Chengalur \& Kanekar}{2000}] {chengalur00} Chengalur J.~N., Kanekar N., 2000, MNRAS, 318, 303

\bibitem[\protect\citeauthoryear{Conselice, Blackburne, \& Papovich}{2005}] {conselice05} Conselice C.~J., Blackburne J.~A., Papovich C., 2005, ApJ, 620, 564
 
\bibitem[\protect\citeauthoryear{Deboer et al. }{2009}] {deboer09} Deboer, D.R., et al., 2009, IEEE Proceedings, 97, 1507 

\bibitem[\protect\citeauthoryear{Erkal, Gnedin, \& Kravtsov}{2012}] {erkal12} Erkal D., Gnedin N.~Y., Kravtsov A.~V., 2012, arXiv, arXiv:1201.3653 

\bibitem[\protect\citeauthoryear{Haehnelt, Steinmetz, \& Rauch}{1998}]{haehnelt98} Haehnelt M.~G., Steinmetz M., Rauch M., 1998, ApJ, 495, 647 

\bibitem[\protect\citeauthoryear{Jonas J. L.}{2009}]{jonas09} Jonas, J.L., 2009, IEEE Proceedings, 97, 1522

\bibitem[\protect\citeauthoryear{Kanekar \& Chengalur}{2003}] {kanekar03} Kanekar N., Chengalur J.~N., 2003, A\&A, 399, 857 

\bibitem[\protect\citeauthoryear{Kanekar et al.}{2009}] {kanekar09} Kanekar N., Smette A., Briggs F.~H., Chengalur J.~N., 2009, ApJ, 705, L40

\bibitem[\protect\citeauthoryear{Karachentsev et al.} {2004}] {kk04} Karachentsev I.~D., Karachentseva V.~E., Huchtmeier W.~K., Makarov D.~I., 2004, AJ, 127, 2031 


\bibitem[\protect\citeauthoryear{Kanekar \& Chengalur}{2003}]{ kanekar03} Kanekar N., Chengalur J.~N., 2003, A\&A, 399, 857 

\bibitem[\protect\citeauthoryear{Kulkarni et al.}{2012}] {kulkarni12} Kulkarni V.~P., Meiring J., Som D., P{\'e}roux C., York D.~G., Khare P., Lauroesch J.~T., 2012, ApJ, 749, 176 


%\bibitem[\protect\citeauthoryear{Lah et al.}{2007}] {lah07} 
%Lah P., et al., 2007, MNRAS, 376, 1357 

\bibitem[\protect\citeauthoryear{Martin et al.}{2010}] {martin10} 
Martin A.~M., Papastergis E., Giovanelli R., Haynes M.~P., Springob C.~M., Stierwalt S., 2010, ApJ, 723, 1359 

\bibitem[\protect\citeauthoryear{McKee \& Krumholz}{2010}]{mckee10} McKee C.~F., Krumholz M.~R., 2010, ApJ, 709, 308 
\bibitem[\protect\citeauthoryear{Milgrom}{1988}] {milgrom88} Milgrom M., 1988, A\&A, 202, L9 

\bibitem[\protect\citeauthoryear{Noterdaeme et 
al.}{2012}]{noterdaeme12a} Noterdaeme P., et al., 2012, arXiv, 
arXiv:1210.1213 

\bibitem[\protect\citeauthoryear{Noterdaeme et al.}{2012}]{noterdaeme12} 
Noterdaeme P., et al., 2012, A\&A, 540, A63 

\bibitem[\protect\citeauthoryear{Noterdaeme et al.}{2009}] {noterdaeme09} Noterdaeme P., Petitjean P., Ledoux C., Srianand R., 2009, A\&A, 505, 1087 

\bibitem[\protect\citeauthoryear{Patra et al.}{2012}] {patra12} Patra N., et al., 2012 (in preparation).

\bibitem[\protect\citeauthoryear{Pontzen et al.} {2008}]{pontzen08} Pontzen A., et al., 2008, MNRAS, 390, 1349 

%\bibitem[\protect\citeauthoryear{Pontzen et al.}{2010}] {pontzen10} %Pontzen A., et al., 2010, MNRAS, 402, 1523 

\bibitem[\protect\citeauthoryear{Prochaska \& Wolfe}{1997}] {prochaska97} Prochaska J.~X., Wolfe A.~M., 1997, ApJ, 487, 73 

\bibitem[\protect\citeauthoryear{Prochaska et al.}{2007}]{prochaska07} Prochaska J.~X., Chen H.-W., Dessauges-Zavadsky M., Bloom J.~S., 2007, ApJ, 666, 267 

\bibitem[\protect\citeauthoryear{Prochaska \& Wolfe}{2009}] {prochaska09} Prochaska J.~X., Wolfe A.~M., 2009, ApJ, 696, 1543 

\bibitem[\protect\citeauthoryear{Rao \& Briggs}{1993}] {rao93} Rao S., Briggs F., 1993, ApJ, 419, 515 

%\bibitem[\protect\citeauthoryear{Razoumov et 
%al.}{2006}]{razoumov06} Razoumov A.~O., Norman M.~L., Prochaska 
%J.~X., Wolfe A.~M., 2006, ApJ, 645, 55 

%\bibitem[\protect\citeauthoryear{Roychowdhury et al.}{2009}] %{roychowdhury09} Roychowdhury S., Chengalur J.~N., Begum A., %Karachentsev I.~D., 2009, MNRAS, 397, 1435 

%\bibitem[\protect\citeauthoryear{Roychowdhury et al.}{2011}]{roychowdhury11} Roychowdhury S., Chengalur J.~N., Kaisin S.~S., Begum A., Karachentsev I.~D., 2011, MNRAS, 414, L55 

\bibitem[\protect\citeauthoryear{Roychowdhury et al.} {2010}] {roychowdhury10} Roychowdhury S., Chengalur J.~N., Begum 
A., Karachentsev I.~D., 2010, MNRAS, 404, L60 

\bibitem[\protect\citeauthoryear{Schaye}{2001}]{schaye01} Schaye 
J., 2001, ApJ, 562, L95 

\bibitem[\protect\citeauthoryear{Swarup et al.}{1991}] {swarup91} 
Swarup G., Ananthakrishnan S., Kapahi V.~K., Rao A.~P., Subrahmanya C.~R., Kulkarni V.~K., 1991, CuSc, 60, 95 

%\bibitem[\protect\citeauthoryear{Watson et al.}{2006}] {watson06} 
%Watson D., et al., 2006, ApJ, 652, 1011 

\bibitem[\protect\citeauthoryear{Verheijen et al.}{2008}]{verheijen08} 
  Verheijen M.~A.~W., Oosterloo T.~A., van Cappellen W.~A., Bakker L., 
  Ivashina M.~V., van der Hulst J.~M., 2008, AIPC, 1035, 265 

%\bibitem[\protect\citeauthoryear{Verheijen et al.}{2008}] {verheijen08} Verheijen, M. A. W., 2008, AIP Proceedings, 1035, 265

%\bibitem[\protect\citeauthoryear{Warren et al.}{2012}]{warren12} 
%Warren S.~R., et al., 2012, ApJ, 757, 84 

\bibitem[\protect\citeauthoryear{Welty, Xue, \& Wong}{2012}] {welty12} Welty D.~E., Xue R., Wong T., 2012, ApJ, 745, 173 

\bibitem[\protect\citeauthoryear{Wolfe \& Davis}{1979}] {wolfe79} Wolfe A.~M., Davis M.~M., 1979, AJ, 84, 699 

\bibitem[\protect\citeauthoryear{Wolfe et al.} {1986}] {wolfe86} 
  Wolfe A.~M., Turnshek D.~A., Smith H.~E., Cohen R.~D., 1986, ApJS,    61, 249 

\bibitem[\protect\citeauthoryear{Wolfe, Gawiser, \& Prochaska}{2005}] {wolfe05} Wolfe A.~M., Gawiser E., Prochaska J.~X., 2005, ARA\&A, 43, 861 

%\bibitem[\protect\citeauthoryear{Young \& Lo}{1997}]{young97} Young L.~M., Lo K.~Y., 1997, ApJ, 490, 710 

\bibitem[\protect\citeauthoryear{Zwaan et al.}{2003}] {zwaan03} 
Zwaan M.~A., et al., 2003, AJ, 125, 2842 

\bibitem[\protect\citeauthoryear{Zwaan et al.}{2005}] {zwaan05} 
Zwaan M.~A., van der Hulst J.~M., Briggs F.~H., Verheijen M.~A.~W., 
Ryan-Weber E.~V., 2005, MNRAS, 364, 1467 


\end{thebibliography}
\end {document}